# FIRST HIGH GRADIENT TEST RESULTS OF A DRESSED 325 MHZ SUPERCONDUCTING SINGLE SPOKE RESONATOR AT FERMILAB

R. C. Webber, T. Khabiboulline, R. Madrak, T. Nicol, L. Ristori, W. Soyars, R. Wagner
FNAL*, Batavia, IL 60510, U.S.A.

*Abstract*

A new superconducting RF cavity test facility has been commissioned at Fermilab in conjunction with first tests of a 325 MHz, β = 0.22 superconducting single-spoke cavity dressed with a helium jacket and prototype tuner. The facility is described and results of full gradient, CW cavity tests with a high $Q_{ext}$ drive coupler are reported. Sensitivities to Q disease and externally applied magnetic fields were investigated. Results are compared to bare cavity results obtained prior to hydrogen degassing and welding into the helium jacket.

## INTRODUCTION

The Fermilab High Intensity Neutrino Source (HINS) program has been developing 325 MHz superconducting spoke resonator accelerating cavities since 2006. Two β = 0.22, single-spoke resonator (SSR1) cavities have been fabricated and both bare cavities have been tested [1][2]. The first cavity, SSR1-01, now dressed with a helium jacket and a prototype tuner, has been tested in conjunction with commissioning the recently completed 325 MHz Superconducting Spoke Cavity Test Facility.

## TEST FACILITY DESCRIPTION

The 325 MHz Spoke Cavity Test Facility (SCTF) in the Fermilab Meson Detector Building (MDB) provides cryogenic fluids, a cryostat, RF power, controls, and data acquisition software designed to support spoke cavity testing and development.

### *Cryogenic Systems*

The cryostat is a 1.44 m OD, 1.1 m inside-length stainless steel vessel containing a warm magnetic shield, an 80 K thermal shield, a cavity support system, and a RF input coupler port. It is capable of cooling any of the test configurations to 4.5 K. A more detailed description of the cryostat and its related components can be found in [3]. Figure 1 is a picture of the cryostat inside the test facility.

Cryogenic fluids are provided by the Meson area Cryogenic Test Facility [4][5]. The helium pressure at the cavity vessel is unregulated, dependent on the pressure drop of the saturated vapor return flow through a 250 m transfer line and a cryogenic heat exchanger into the refrigerator compressor suction. The vessel pressure is approximately 130 kPa with no RF heat. When RF power is introduced, the helium flow and, consequently, the cavity vessel pressure increase as a function of the heating. The maximum allowable working pressure of the cavity vessel is 170 kPa; this becomes the limiting factor for cooling capacity.

In the present configuration, cavity cool-down from room temperature to 5 K takes 15 hours. About three hours are required to transit the 150-70 K Q disease sensitive zone.

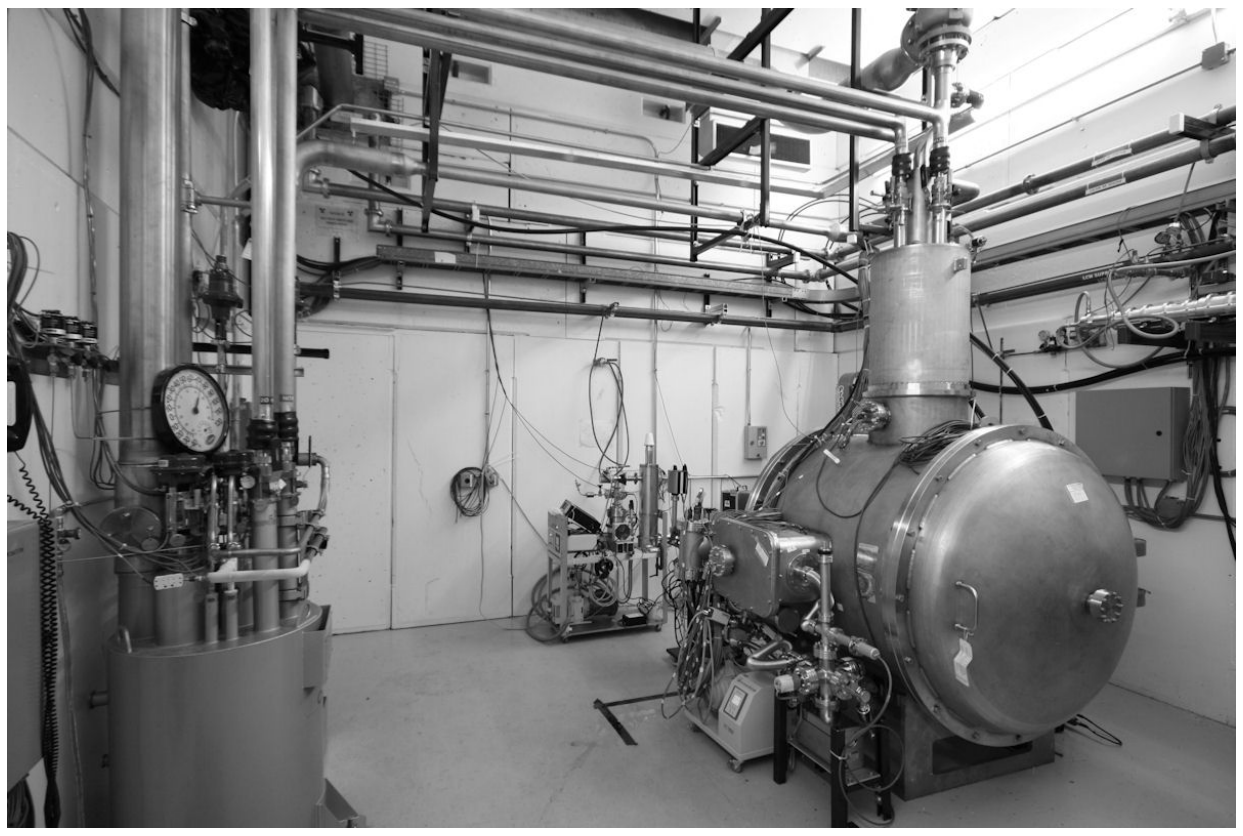

Figure 1. Photo of the Cavity Test Facility Cryostat

### *RF Systems*

The SCTF facility offers both a CW 200 W, 325 MHz power source and a pulsed 250 kW, 325 MHz source capable of pulse lengths up to 3 ms at a 1% duty factor. With a high $Q_{ext}$ drive antenna, the 200 W source is sufficient to drive a SSR1 cavity beyond its design field. The pulsed RF source will be used for dynamic Lorentz force cavity detuning tests and power coupler tests.

A modern digital low level RF system (LLRF) [6] with GUI interface screens controls either power source. The LLRF offers a 20 kHz bandwidth feedback loop for tracking the cavity resonant frequency. The SCTF LLRF and slow controls interface to Fermilab's ACNET control system for data logging and other standard services.

## SSR1-01 HISTORY

Two SSR1 prototypes were fabricated to Fermilab specifications: SSR1-01 by E. Zanon (Schio, Italy), and SSR1-02 by Roark (Brownsburg, Indiana, USA). Both bare cavities have been tested in the Fermilab Vertical Test Stand (VTS).

SSR1-01 was cooled down and tested several times at 4.4 and 2.0 K in VTS. After the second cool down, $Q_0$ was a factor of two worse at 2 K for low $E_{acc}$ (~2 MV/m) than after the first cool down. The Q-slope, $Q_0$ vs. $E_{acc}$, was also much steeper. These symptoms are indicative of Q disease despite the fast, 20-minute, cool down time at

#webber@fnal.gov

VTS through the 150-70 K sensitive region. An explicit test for Q disease at 4.4 K was performed by holding SSR1-01 at 100 K for seven hours during its final cool down in VTS. Figure 2 shows the results. $Q_0$ dropped by about a factor of eight over most of the $E_{acc}$ range.

Time constraints prevented further tests of SSR1-01 in the VTS. It was sent to JLAB for hydrogen degassing (600 °C vacuum bake for 10 hours). The cavity was then tuned at room temperature to achieve the required operating frequency at 4.4 K. The stainless steel helium jacket was welded to the cavity, and a 20-minute flash, buffered chemical process and a high-pressure rinse (HPR) were performed on it at the ANL G150 facility.

## CAVITY TESTS AND RESULTS AT SCTF

The jacketed SSR1-01 cavity was dressed with prototype tuners and a high, ~1.5e8, $Q_{ext}$ drive antenna for CW testing in SCTF. Upon initial cool down to ~4.5 K, the cavity resonant frequency was within 20 kHz of the expected value. The cavity and stainless steel vessel behaved as expected during cool down.

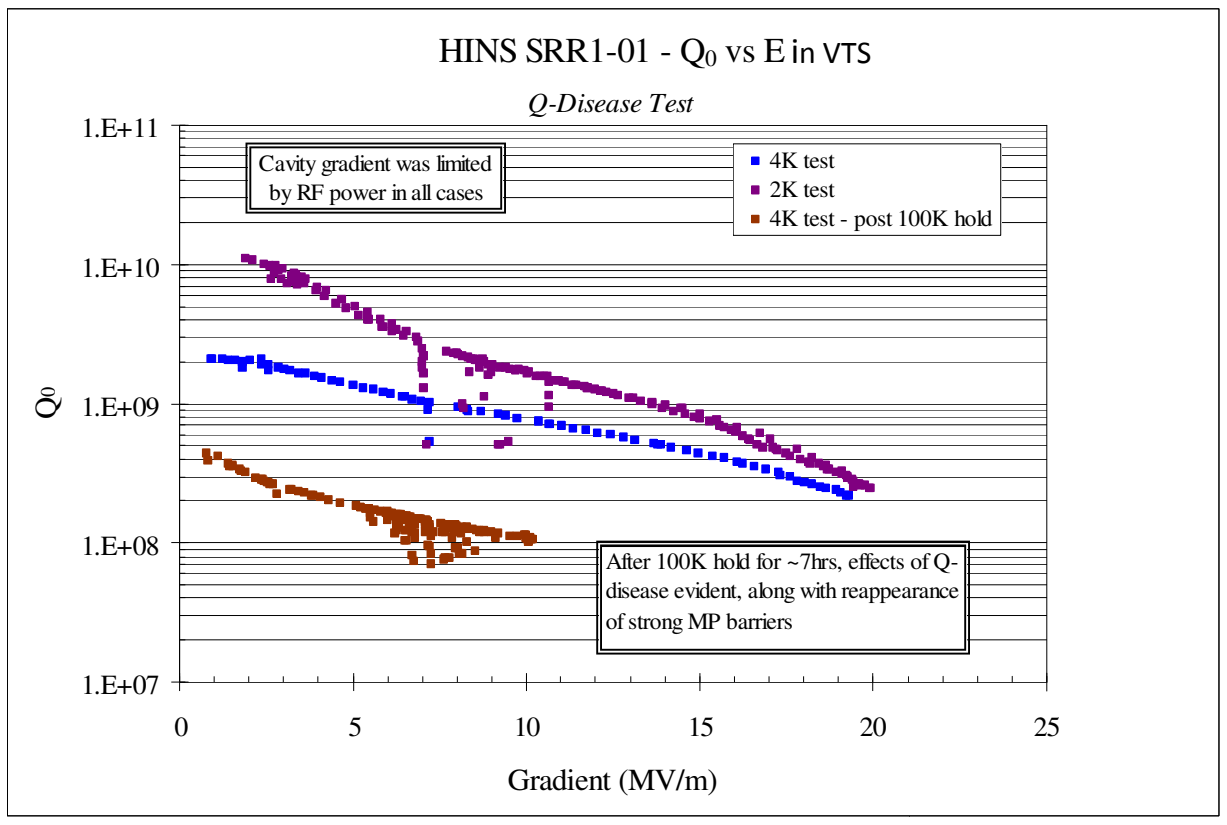

Figure 2. $Q_0$ vs. $E_{acc}$ results results of bare SSR1-01 in VTS prior to degassing and dressing. Upper curve at 2.0K, middle curve at 4.4K, lower curve at 4.4K after attempting to induce Q disease symptoms

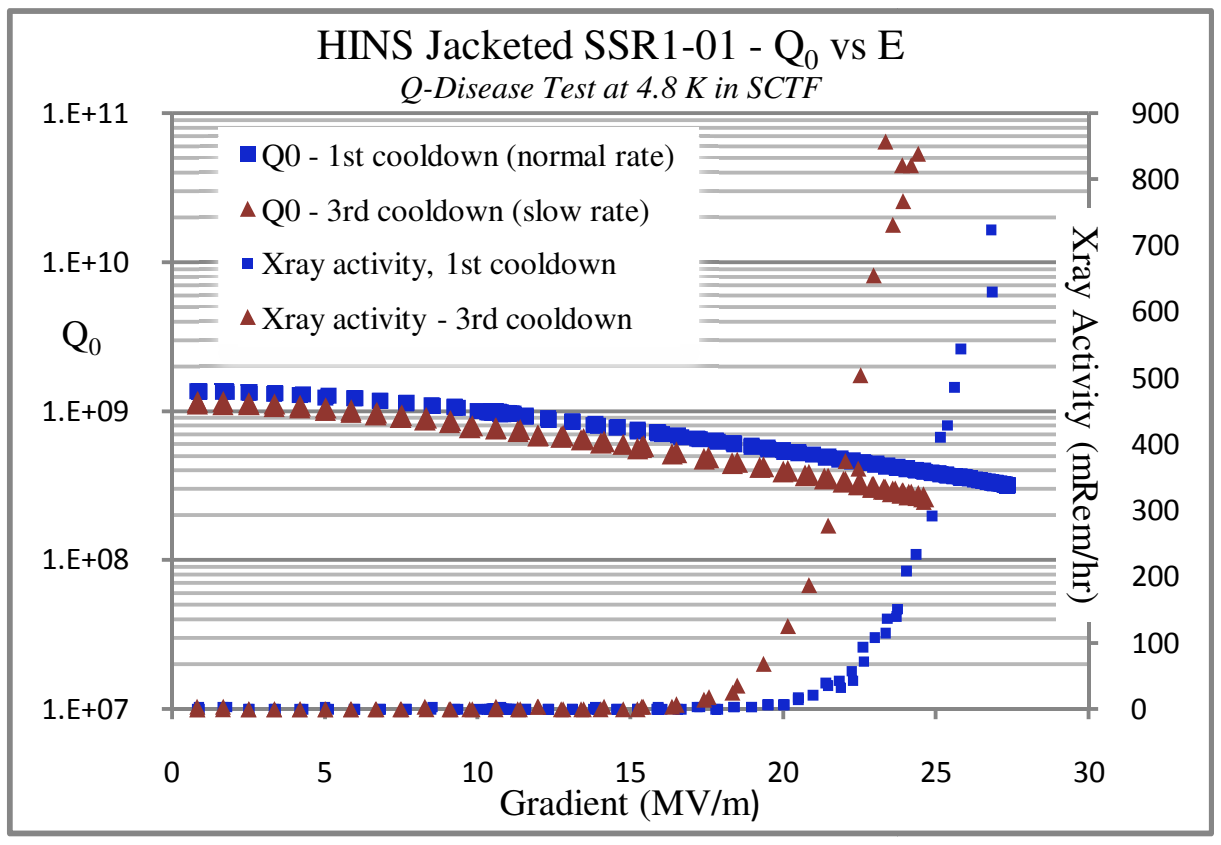

Figure 3. $Q_0$ vs. $E_{acc}$ results and x-ray activity of SSR1-01, degassed and dressed, in SCTF before (upper) and after (lower) the attempt to induce Q disease symptoms

The cavity tuners were engaged to apply an inward force of 6 kN on the beam pipe on each end of the cavity. In this state, the resonant frequency sensitivity to helium pressure was measured to be -145 ± 15 Hz/torr, in fair agreement with -210 Hz/torr predicted by simulation. The static Lorentz force detuning coefficient was found to be -1.5 ± 0.5 Hz/(MV/m)$^2$, considerably lower than -3.8 Hz/(MV/m)$^2$ predicted by simulation.

RF conditioning to clean up multipactoring and attain high field operation required ~10 hours. The ability to switch the CW RF amplitude between relatively low and high power with a variable duty cycle and a few second period proved useful. Operating in this mode was necessary for high $E_{acc}$ operation, where x-ray emissions increase power loss. Drive power was switched between low and high values with suitable duty factor to keep the average loss below the 35 watt limit imposed by system pressure constraints.

In this SCTF test, SSR1-01 reached a quench-limited maximum $E_{acc}$ (voltage over effective distance between beam pipe irises) of 27 MV/m. Figure 3 shows a $Q_0$ vs. $E_{acc}$ scan for SSR1-01 at 4.7 K during its first SCTF cool down. For comparison, $Q_0$ vs. $E_{acc}$ scans at 2.0 K and 4.4 K for the bare SSR1-01 in VTS, after several typical VTS speed cool downs, are shown in Figure 2. The 27 MV/m achieved by SSR1-01 at SCTF is considerably higher than what was achieved at 4.4 K in VTS. Therefore, SSR1-01 performed well in its first cold test after the hydrogen degassing, room temperature tuning, jacketing, flash BCP and HPR.

To test for Q disease after the hydrogen degassing, a warm up to room temperature was followed by a second cool down with similar cooling rate, ~3 hours, through the sensitive region. The subsequent $Q_0$ .vs. $E_{acc}$ at ~4.7 K was consistent with that for the first cool down within the 5% measurement uncertainty. A third, slower cool down was done spending 11 hours in the sensitive region. Figure 3 compares the $Q_0$ vs. $E_{acc}$ results. Q disease is still evident, but much less than before the hydrogen degassing. At 10 MV/m, there is 20% decrease in $Q_0$ in this test compared to 87% in the VTS slow cool down test (see Figure 2). SSR1-01 was instrumented with x-ray detectors both inside and outside of the SCTF cryostat. Figure 3 shows correlations between x-ray activity and cavity field.

These tests indicate no significant degradation in $Q_0$ at ~4.5 K if cool down times through the sensitive region are ~3 hours or less. Extrapolating these measurements to the effect at 2.0 K is problematic, since the 5% uncertainty at 4.5 K translates to a ~50% uncertainty at 2.0 K. It is unlikely that 11 hours in the sensitive region would be acceptable at 2.0 K.

SSR1 applications in the proposed Fermilab Project X [7][8] anticipate beam focusing solenoid magnets in close proximity to the cavities. Susceptibility of the SSR1 design to magnet fields was tested. Two weak solenoid coils were installed outside each cavity end wall. They were aligned azimuthally with the spoke where the surface RF magnetic field is highest. These coils give a 2 G maximum magnetic field at the cavity end-wall per ampere of current. The maximum fringe field of an actual beam focusing solenoid is expected to be 0.1 G at the cavity.

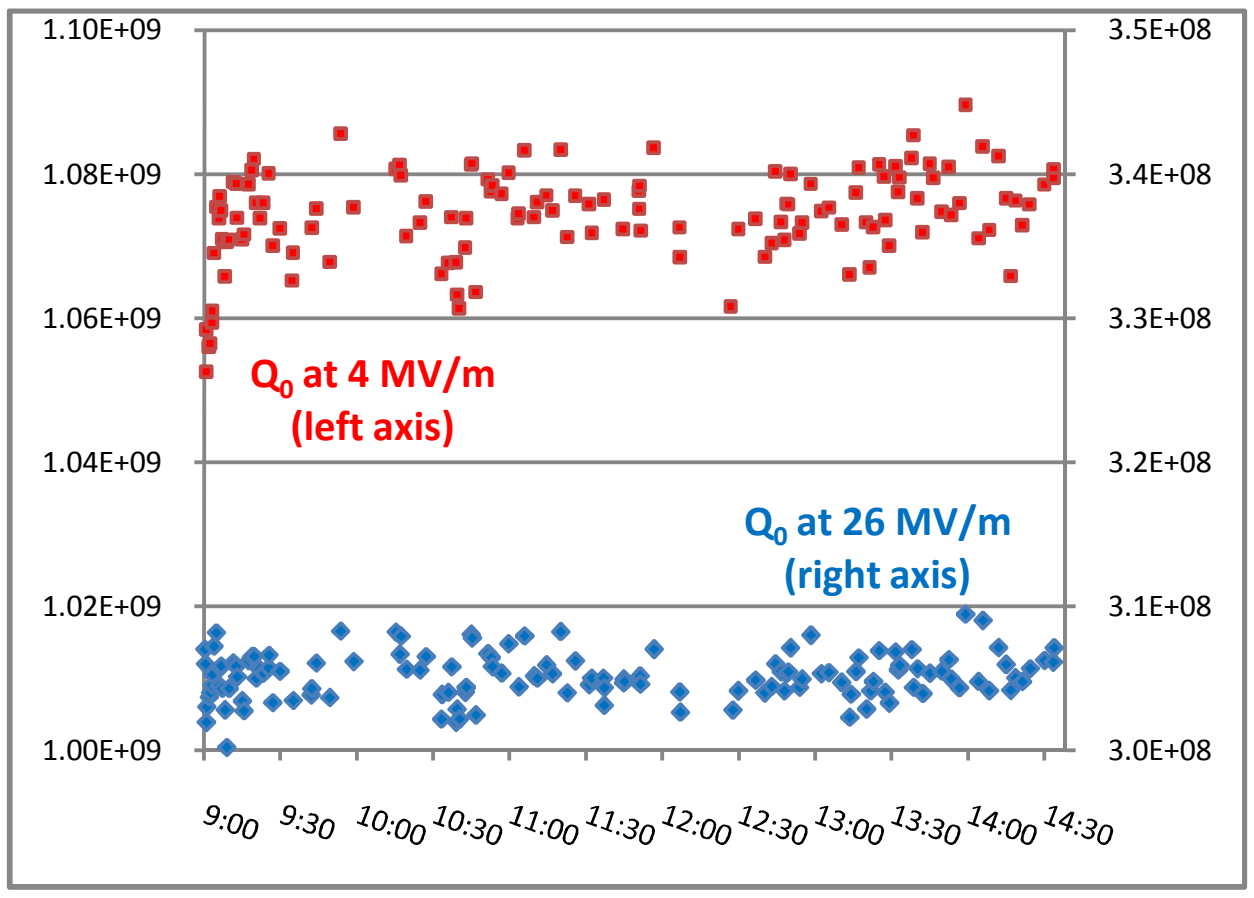

Figure 4. SSR1-01 $Q_0$ at low and high $E_{acc}$ during period of multiple quenches in presence of 8-10 G magnetic field at the cavity end-walls

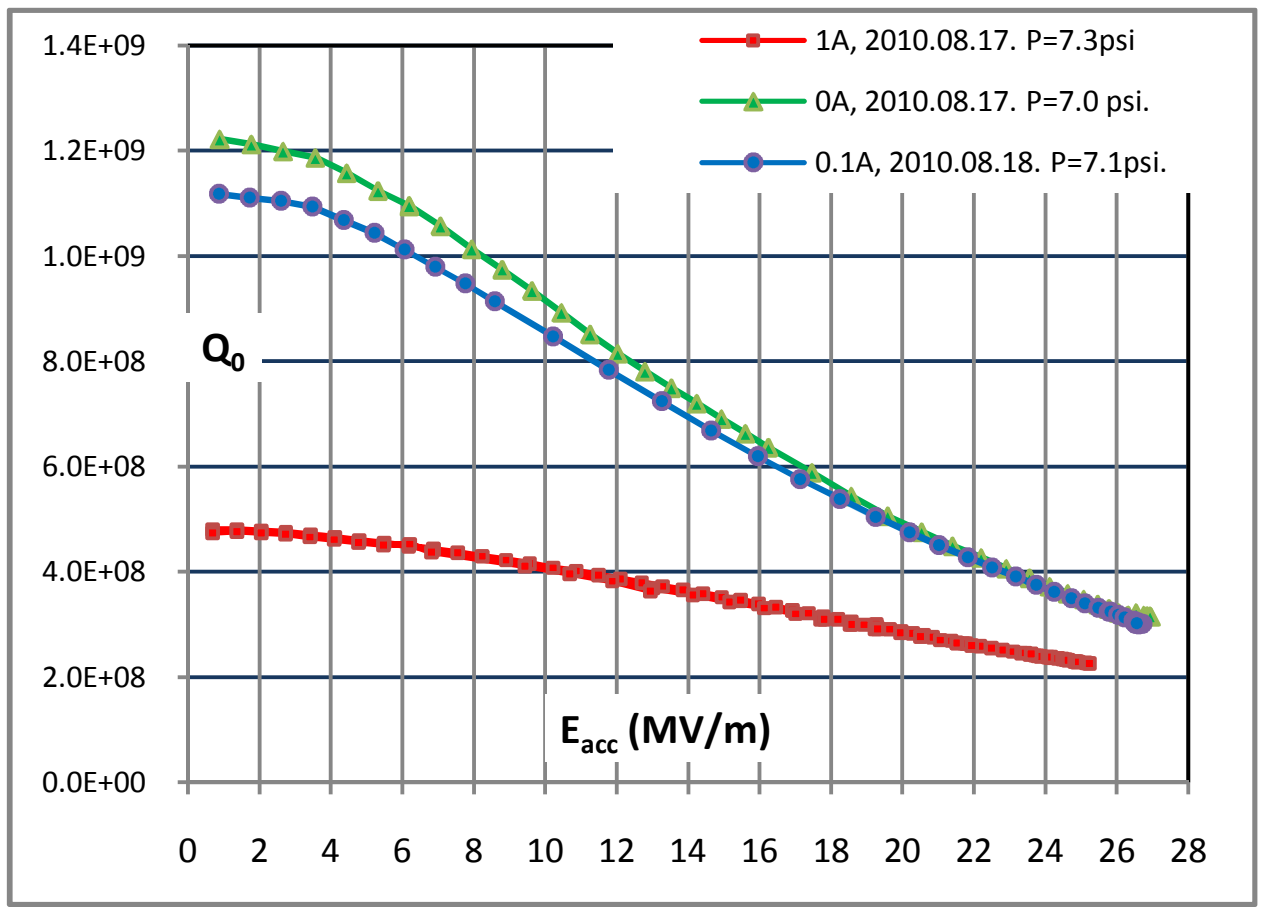

Figure 5. SSR1-01 $Q_0$ degradation due to magnetic field captured during cooling down. (1 A (red) is 2 G field)

One test involved applying a magnetic field to the cavity already in a superconducting state, and then watching the effect on $Q_0$ of repeatedly quenching the cavity with RF in the presence of this field. With 4-5A magnet current (8-10 G at cavity end-wall), no degradation of $Q_0$ at either low or high cavity gradient was observed after 5000 quenches. The quenches are believed to occur at locations on the cavity spoke where the surface RF magnetic field is highest. Since there was no effect on $Q_0$, the conclusion is that superconducting cavity end-walls effectively shield the spokes and very little magnetic field is trapped during quench. Figure 4 shows the insensitivity of $Q_0$ vs. time (~number of quenches) for this test that was done at 4.7 K.

The second test involved cooling the cavity from the normal to the superconducting state in the presence of a magnetic field. At ~4.7 K, the effect of a 0.2 G field is about a 10% reduction of $Q_0$ at low gradient and less at higher gradient. A 2 G field causes nearly a factor of three deterioration of the low field $Q_0$. Figure 5 shows these results.

## FUTURE SCTF AND SSR1 PLANS

Soon a high power RF input coupler, suitable for Project X application, will be installed on SSR1-01 and high, pulsed power RF tests will ensue at SCTF. While pulsed operation at ~4.7 K is no longer directly relevant to Project X, this will be the first cold test of the cavity and power coupler assembly. Dynamic Lorentz force detuning and compensation using piezo actuators will also be studied.

Plans are underway to extend SCTF operation to 2 K in FY11. Similar measurements as reported here will be repeated at 2 K and the study of microphonics and compensation with the fast piezo tuners at that temperature will be of great interest. SSR1-02 and twelve new SSR1 cavities, expected to arrive at Fermilab in 2011, are ultimately destined for testing at SCTF.